\documentclass[conference]{IEEEtran}

\usepackage{amsmath,amssymb,amsfonts}
\usepackage{algorithmic}
\usepackage{graphicx}
\usepackage{subcaption}
\usepackage{textcomp}
\usepackage{booktabs}
\usepackage{float}
\usepackage{todonotes}
\usepackage[hidelinks]{hyperref}
\usepackage{cite}
\usepackage{url}

\usepackage{tikz}
\usetikzlibrary{arrows.meta, positioning, shapes.geometric}
\usepackage{pgfplots}
\usepackage{pgfplotstable}
\pgfplotsset{compat=1.18}
\usepackage{smartdiagram}

\usepackage{acronym}
\newacro{fdna}[FDNA]{FPGA dataflow neural network accelerator}
\newacro{dnn}[DNN]{deep neural network}
\newacro{qnn}[QNN]{quantized neural network}
\newacro{mac}[MAC]{multiply-accumulate}
\newacro{ptq}[PTQ]{post-training quantization}
\newacro{qat}[QAT]{quantization-aware training}
\newacro{mpe}[MPE]{matrix of processing engines}
\newacro{llm}[LLM]{large language model}
\newacro{hls}[HLS]{high-level synthesis}
\newacro{pot}[PoT]{power-of-two}
\newacro{lsq}[LSQ]{learned step size quantization}
\newacro{mcu}[MCU]{microcontroller}
\newacro{fpu}[FPU]{floating-point unit}
\newacro{ste}[STE]{straight-through estimator}
\newacro{bn}[BN]{batch normalization}
\newacro{msb}[MSBs]{most significant bits}
\newacro{lsb}[LSBs]{least significant bits}
\newacro{simd}[SIMD]{single instruction multiple data}
\newacro{rn8}[RN-8]{ResNet-8}
\newacro{rn18}[RN-18]{ResNet-18}
\newacro{bs}[BS]{bit-serial}
\newacro{alu}[ALU]{arithmetic logic unit}
\newacro{sp-nets}[SP-nets]{switchable-precision networks}
\newacro{isa}[ISA]{instruction-set architecture}
\newacro{ml}[ML]{machine learning}
\newacro{gpu}[GPU]{graphics processing unit}
\newacro{fpga}[FPGA]{field-pro\-gram\-ma\-ble gate array}
\newacro{nn}[NN]{neural network}
\newacro{pl}[PL]{programmable logic}
\newacro{ps}[PS]{processing system}
\newacro{ocm}[OCM]{on-chip memory}
\newacro{pfn}[PFN]{pillar feature net}
\newacro{mfcc}[MFCC]{mel-frequency cepstral coefficient}
\newacro{kws}[KWS]{keyword spotting}
\newacro{asic}[ASIC]{application-specific integrated circuit}
\newacro{mems}[MEMS]{micro-electromechanical systems}
\newacro{pdm}[PDM]{pulse-density modulation}
\newacro{pcm}[PCM]{pulse-code modulation}
\newacro{cic}[CIC]{cascaded-integrator-comb}
\newacro{dsconv}[DSConv]{depthwise separable convolution}
\newacro{osr}[OSR]{oversampling ratio}
\newacro{pe}[PE]{processing element}
\newacro{gscd}[GSCD]{Google Speech Commands Dataset}
\newacro{fifo}[FIFO]{first-in-first-out}
\newacro{bram}[BRAM]{block random access memory}
\newacro{lut}[LUT]{look-up table}
\newacro{dsp}[DSP]{digital signal processing}
\newacro{rtl}[RTL]{register-transfer level}
\newacro{cdc}[CDC]{clock domain crossing}
\newacro{adc}[ADC]{analog-to-digital converter}
\newacro{fir}[FIR]{finite impulse response}
\newacro{ascd}[ASCD]{Accented Speech Command Dataset}
\newacro{mvau}[MVAU]{Matrix-Vector-Activation Unit}
\newacro{fc}[FC]{fully connected}

\newcommand{\hms}[1]{\textcolor{blue}{}}
\newcommand{\cb}[1]{\textcolor{red}{}}
\newcommand{\yu}[1]{\textcolor{green}{}}

\newcommand{\archparam}[1]{\mbox{\large$\mathbf{#1}$}}

\begin{document}

\bstctlcite{IEEEtran:BSTcontrol}

\title{\texttt{\textbf{Tartar}}: A Lightweight Raw-Waveform 
\\ Keyword Spotting Model}

\author{
\IEEEauthorblockN{Charalampos Bezaitis\textsuperscript{1}, Magnus Sj{\"a}lander\textsuperscript{1}, Di Liu\textsuperscript{1}, and Yaman Umuroglu\textsuperscript{1}}
\IEEEauthorblockA{
\textit{Norwegian University of Science and Technology} \\
FirstName.LastName@NTNU.no}
}

\maketitle

\begin{abstract}
Keyword spotting (KWS) is a common always-on, machine-learning workload for edge devices. 
End-to-end models that classify raw waveforms avoid expensive preprocessing, previously shown to be a bottleneck,
but lightweight deployment still requires jointly tuning accuracy against both MAC and non-MAC costs.
We present \texttt{Tartar}, a raw-waveform KWS model that is ``tunable'' with a carefully selected set of architectural parameters, enabling trading accuracy against hardware cost. 
When evaluating \texttt{Tartar}, we find that non-MAC layers significantly impact the hardware cost.
We introduce a lossless compression technique and its hardware implementation that significantly reduces the non-MAC cost.
%
\texttt{Tartar} achieves 95.93\% accuracy on the Google Speech Commands Dataset v2 with an implementation of only 4{,}028 LUTs, 8 DSPs, and 16 BRAMs.
Compared against the closest competing KWS model, \texttt{Tartar} uses $7\times$ fewer LUTs and $10\times$ fewer DSPs, 
while remaining competitive in accuracy.
\end{abstract}

\section{Introduction}
\label{sec:introduction}

\Ac{kws} is an always-on \ac{ml} application that detects one or more keywords in a stream of audio.
The always-on feature enables spoken words to activate further functionality, but requires low-power solutions to avoid high energy usage. 
As \ac{kws} is widely used in many devices, from mobile phones to smart speakers, it is important to be as efficient and low-power as possible.
However, \ac{ml} inference is only one part of the end-to-end \ac{kws} pipeline, and 
the preprocessing stage can have a significant impact on the application's latency, power, and memory usage. 
Therefore, accelerating only the neural network itself offers limited improvements and is subject to Amdahl's law~\cite{amdahl:AFIPS1967}.

\looseness -1
The preprocessing step typically consists of computing \ac{mfcc}~\cite{helloegde:ARXIV2018} features; see \autoref{fig:pipelines}. 
Previous works~\cite{fixedmfcc:AICAS2021, serialMFCC:JSCC2021} have 
proposed efficient implementations of the \ac{mfcc}, but even then
it remains a bottleneck for keyword spotting. 
Fariselli et al. proposed a solution that uses fixed-point operations on a \ac{mcu} platform, where the \ac{mfcc} still accounts for 50\% of the total latency~\cite{fixedmfcc:AICAS2021}.
Similar results are found in \ac{asic} implementations, where the \ac{mfcc} consumes twice as much power as the neural network~\cite{serialMFCC:JSCC2021}.

Some earlier works avoid the \ac{mfcc} altogether and instead use a neural network to learn custom convolution filters that operate directly on the raw \ac{pcm} waveform~\cite{sincnet:SLT2018, leaf:ICLR2021, yadav:INTERSPEECH2022}, also referred to as end-to-end \ac{kws} models.
One such model is SincConv~\cite{sincconv:ICASSP2020}, which achieves state-of-the-art accuracy by training parametrized $\mathrm{sinc}$ filters as the first layer of the network.

\begin{figure*}[t]
    \centering
\input{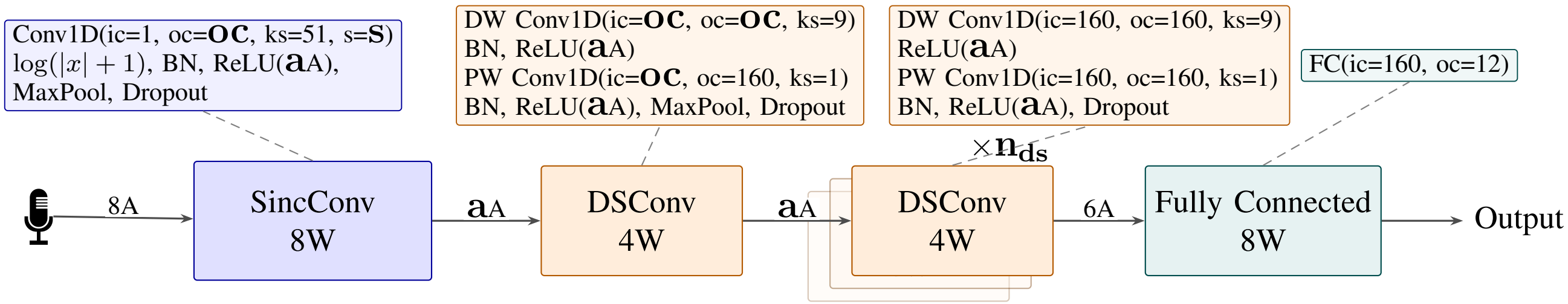}


\def\xmic{0}       
\def\xstep{4}    
\def\xconv{3.0}    
\def\xdsone{6.6}   
\def\xdstwo{10.0}   
\def\xclf{13.4}    
\def\xoutput{16.2} 

\def\xconvd{1.8}
\def\xdsonecall{6.8}
\def\xdstwocall{11.5}
\def\xclfd{15.0}
\def\ycall{1.7}

\begin{tikzpicture}[
  block/.style={
    rounded corners=1pt,
    inner xsep=3pt,
    inner ysep=2pt,
    align=center,
    line width=0.4pt,
    font=\normalsize,
  },
  convblock/.style={block, draw=blue!60!black, fill=blue!12,
    minimum width=26mm, minimum height=13mm},
  dsblock/.style={block, draw=orange!70!black, fill=orange!14,
    minimum width=22mm, minimum height=12mm},
  dsghost/.style={dsblock, opacity=0.28, fill=orange!8, draw=orange!45!black},
  repeatbadge/.style={
    anchor=south east,
    inner sep=1pt,
    fill=white,
    rounded corners=1pt,
  },
  clfblock/.style={block, draw=teal!60!black, fill=teal!10,
    minimum width=20mm, minimum height=12mm},
  small/.style={
    rounded corners=1pt,
    inner xsep=2.5pt,
    inner ysep=1.5pt,
    align=left,
    line width=0.35pt,
    font=\footnotesize,
  },
  convsmall/.style={small, draw=blue!60!black, fill=blue!6},
  dssmall/.style={small, draw=orange!70!black, fill=orange!8},
  dssmalltwo/.style={small, draw=orange!70!black, fill=orange!8},
  clfsmall/.style={small, draw=teal!60!black, fill=teal!6},
  arr/.style={-{Stealth[length=1.4mm, width=1mm]}, semithick, draw=black!70},
  callout/.style={densely dashed, draw=black!50},
  aqlabel/.style={
    inner sep=0.8pt,
    fill=white,
    font=\footnotesize,
  },
]

\node[inner sep=0] (mic) at (\xmic, 0.05)
  {\scalebox{0.55}{\pipelinemic}};
\node[convblock] (conv1) at (\xconv, 0) {%
  SincConv\\
  8W%
};
\node[dsblock] (ds1) at (\xdsone, 0) {%
  DSConv\\
  4W%
};
\def\dsstackdx{-2.4mm}
\def\dsstackdy{-1.4mm}
\node[dsghost] (ds2g2) at ([xshift={2*\dsstackdx}, yshift={2*\dsstackdy}] \xdstwo, 0) {%
  DSConv\\
  4W%
};
\node[dsghost, opacity=0.42] (ds2g1) at ([xshift=\dsstackdx, yshift=\dsstackdy] \xdstwo, 0) {%
  DSConv\\
  4W%
};
\node[dsblock] (ds2) at (\xdstwo, 0) {%
  DSConv\\
  4W%
};
\node[repeatbadge] at (ds2.north east) {{\normalsize$\times$}\archparam{n_{ds}}};
\node[clfblock] (clf) at (\xclf, 0) {%
  Fully Connected\\
  8W%
};
\node[font=\normalsize] (out) at (\xoutput, 0) {Output};

\draw[arr] (mic) -- node[aqlabel, above] {8A} (conv1);
\draw[arr] (conv1) -- node[aqlabel, above] {\archparam{a}A} (ds1);
\draw[arr] (ds1) -- node[aqlabel, above] {\archparam{a}A} (ds2);
\draw[arr] (ds2) -- node[aqlabel, above] {6A} (clf);
\draw[arr] (clf) -- (out);

\node[convsmall] (convd) at (\xconvd, \ycall) {%
  Conv1D(ic=1, oc=\archparam{oc}, ks=51, s=\archparam{s})\\
  $\log(|x|+1)$, BN, ReLU(\archparam{a}A), \\MaxPool, Dropout%
};
\draw[callout] (conv1.north) -- (convd.south);

\node[dssmall] (ds1d) at (\xdsonecall, \ycall) {%
  DW Conv1D(ic=\archparam{oc}, oc=\archparam{oc}, ks=9)\\
  BN, ReLU(\archparam{a}A)\\
  PW Conv1D(ic=\archparam{oc}, oc=160, ks=1)\\
  BN, ReLU(\archparam{a}A), MaxPool, Dropout%
};
\draw[callout] (ds1.north) -- (ds1d.south);

\node[dssmalltwo] (ds2d) at (\xdstwocall, \ycall) {%
  DW Conv1D(ic=160, oc=160, ks=9)\\
  ReLU(\archparam{a}A)\\
  PW Conv1D(ic=160, oc=160, ks=1)\\
  BN, ReLU(\archparam{a}A), Dropout%
};
\draw[callout] (ds2.north) -- (ds2d.south);

\node[clfsmall] (clfd) at (\xclfd, \ycall) {%
  FC(ic=160, oc=12)%
};
\draw[callout] (clf.north) -- (clfd.south);

\end{tikzpicture}
    \caption{ 
    The \texttt{Tartar} \ac{kws} model architecture.
    Bold values denote tunable architectural parameters: output channels \archparam{oc} and stride \archparam{s} of the first convolution; \archparam{n_{ds}} the number of repeated \ac{dsconv} blocks after the first; and \archparam{a} the activation bit-width. 6A indicates 6-bit activation quantization and 8W indicates 8-bit weight quantization.}
    \label{fig:the_model}
\end{figure*}

This work presents \texttt{Tartar}, an end-to-end \ac{kws} machine-learn\-ing model that enables lightweight implementations. 
%
%
\texttt{Tartar} is tunable with a small set of key architectural parameters; see \autoref{fig:the_model}, enabling accuracy and hardware cost to be traded for different always-on targets.
With a systematic approach, we evaluate the impact of the parameters on \texttt{Tartar}'s resource cost
in terms of both \ac{mac} and non-\ac{mac} operations, as well as memory usage.
Based on this approach, \texttt{Tartar} can be optimized for a range of throughputs, hardware resources, and accuracies.

Our main contributions are:
\begin{itemize}
\item A tunable, fully quantized end-to-end KWS model operating on raw audio waveforms.
\item A systematic exploration across a set of key architectural parameters that showcase their importance for 
lightweight \ac{kws} implementations.
\item A novel delta-compression scheme for lossless compression of activation thresholds and its hardware implementation, reducing non-\ac{mac} resource cost by up to $4\times$. 
\item A comparison against the closest competing FPGA KWS model~\cite{leekws:ICASSP2026} showing $7\times$ fewer LUTs and $10\times$ fewer DSPs, 
while retaining competitive accuracy.
\end{itemize}

\section{\texttt{Tartar}: A Raw-Waveform KWS Model}
\label{sec:tartar}

\texttt{Tartar} operates directly on raw \ac{pcm} input, analogous to culinary tartare being served raw. 
By jointly optimizing accuracy and resource consumption in a single unified model, \texttt{Tartar} produces a lightweight design tailored for resource-constrained deployment (\autoref{sec:every_bit_operation_matters}).
The model follows a three-stage layout: %
\textit{(i)}~a SincConv layer, %
\textit{(ii)}~a sequence of \acf{dsconv} layers, and %
\textit{(iii)}~a \ac{fc} output layer.
\texttt{Tartar} uses a similar layout to SincConv~\cite{sincconv:ICASSP2020}, adding a small final \ac{fc} layer while reducing the SincConv layer's memory and computational cost and tuning the total number of layers.

On \ac{gscd} v2, each example is a one-second-long \ac{pcm} waveform sampled at 16 kHz.
The first learned layer, the SincConv, maps the 16{,}000 \ac{pcm} signed 8-bit samples to a smaller set of feature maps, which the \acp{dsconv} process. 
Finally, a fully connected layer with 12 output channels produces the final response.
The whole model architecture is shown in \autoref{fig:the_model}.

\subsection{SincConv}
\label{sec:quantizing_sincnet}

\texttt{Tartar} employs $\mathrm{sinc}$ functions in the first layer that operate directly on the \ac{pcm} waveform.
In the case of audio, a $\mathrm{sinc}$ function in the time domain is a rectangular function in the frequency domain and acts as a band-pass filter.
SincNet~\cite{sincnet:SLT2018} uses this idea to learn the filter coefficients $f_1$ and $f_2$ through backpropagation.
\begin{equation}
    g(n,f_{1},f_{2}) = 2f_{2}sinc\!\left(2\pi f_{2}n\right) - 2f_{1}sinc\!\left(2\pi f_{1}n\right)
\end{equation}
Depending on the number of filters $N$, which is selected by the user, 
each filter $\mathbf{h}^{(i)}$ uses its own learned cut-off frequencies $f_{1}^{(i)}$ and $f_{2}^{(i)}$, while its dimension is $D$:
\begin{equation}
\label{eq:filters}
\mathbf{H}=[\mathbf{h}^{(1)};\ldots;\mathbf{h}^{(N)}]
\quad
\mathbf{h}^{(i)}=[h^{(i)}_{0},\ldots,h^{(i)}_{D-1}]
\end{equation}
In convolution terms, the number of filters $N$ is the number of output channels and the dimension $D$ is the kernel size. 
The filters are transposed to be the weights of a 1D convolution.
The output of the convolution is then passed through log-compression (i.e., $\log(\lvert x \rvert+1)$)~\cite{sincconv:ICASSP2020}. 

The SincConv filters of \autoref{eq:filters} are first realized as a 1D convolution with floating-point weights.
The floating-point weights are then passed through a trainable quantization node. 
The trainable quantization node and the sinc generation from the learned frequency ranges stay in the computational graph during \ac{qat}.
After training, the weights are transformed and kept as 8-bit signed weights of an ordinary 1D convolution.
Further reducing the weights' bit-width causes a significant loss in accuracy.
This 1D convolution serves as the first layer of \texttt{Tartar}.

\subsection{Depthwise Separable Convolution}
\label{sec:dsconv}

\texttt{Tartar} employs \acfp{dsconv}, which were first popularized for efficient mobile vision~\cite{mobilenet_v1:ARXIV2017} and later adopted for audio keyword spotting~\cite{helloegde:ARXIV2018}.
A \ac{dsconv} combines a depthwise convolution, which applies a kernel to each input channel, with a pointwise convolution using a small kernel.
For this audio deployment, both convolutions are 1D.
Each \ac{dsconv} block is trained with \ac{qat} to 4-bit weights, with activations quantized after both the depthwise and pointwise convolutions.
\Ac{ptq} is avoided as this type of low-parameter convolution is difficult to calibrate after training, especially for low bit-widths~\cite{mobilenet_bad_quant:CVPR2021}.

\subsection{Fully Connected Layer}
\label{sec:fclayer}

Following common practice in low bit-width deep neural networks~\cite{a2q:ICCV2023}, the final fully connected layer is kept at higher precision than the \ac{dsconv} blocks: 8-bit signed weights with 6-bit activations.

\section{Every Bit and Bit-Operation Count}
\label{sec:every_bit_operation_matters}

\Acf{kws} is a utility function that, in principle, can wake up any system with a microphone, provided its hardware implementation has negligible area and power requirements.
To achieve such a low-cost and low-power design, every bit and operation count.

\subsection{Lightweight Raw-Waveform KWS} 
\label{sec:model_parametrization_optimization}

Techniques such as low-precision quantization reduce the cost of performing \ac{mac} operations, but they can increase the relative overhead of non-\ac{mac} operations.
These costs come from floating-point scaling factors and normalization layers. 
In models with larger convolutions, this overhead is often negligible.
In a lightweight design like \texttt{Tartar}, the non-\ac{mac} operations can drive resource cost, so we parametrize the activation bit-width \archparam{a}, which controls the non-\ac{mac} cost.
An expensive first layer, such as SincConv, influences the hardware mapping of the entire pipeline disproportionately, which is why we select the number of output channels \archparam{oc} and the stride \archparam{s} as input parameters to the model.
Beyond hardware cost, iterative training shows that these parameters have a significant impact on accuracy, as seen in \autoref{sec:accuracy_exploration}.
A final architectural parameter is the number of \ac{dsconv} blocks \archparam{n_{ds}}, which enables deeper networks with higher accuracy. 
All four parameters, \archparam{oc}, \archparam{s}, \archparam{a}, and \archparam{n_{ds}}, are tunable, as seen in \autoref{fig:the_model}.
%
These architectural parameters determine hardware resource usage as seen in \autoref{sec:conv_resource_cost} and \autoref{sec:layer_tail_resource_cost},
while they also determine the accuracy of the model as seen in \autoref{sec:accuracy_exploration}.

We keep the SincConv kernel size fixed at $ks=51$ as smaller kernels lose accuracy, while larger ones ($71$--$151$) raise the expensive SincConv's cost with no accuracy gain.
We keep the \ac{dsconv} architectural parameters fixed, following previous work~\cite{sincconv:ICASSP2020}, to keep the search space tractable.

\subsection{Convolutional Resource Cost}
\label{sec:conv_resource_cost}

The following sections describe how the \archparam{oc}, \archparam{s}, \archparam{a}, and \archparam{n_{ds}} parameters impact resource usage of the convolutions in terms of compute and memory cost.

\subsubsection{Architectural Parameter Impact on Bit-Operations}
\label{sec:param_bitops}

Since the cost of a 4-bit \ac{mac} is not the same as an 8-bit \ac{mac}, we adopt the bit-operations metric, which is commonly used to measure \ac{mac} cost in low bit-width \acp{nn}~\cite{bitoperation:TOCS2021} \hms{We claim common, so can we add some more references to show that it is used in more than one article?}.
A 4-bit multiplication performs $4\times4=16$ bit-operations while an 8-bit multiplication performs $8\times8=64$, resulting in a four-fold difference in bit-operations.

\looseness -1
For a typical convolution, the number of \acp{mac} is a function of the input size $n$, output channels $oc$, input channels $ic$, stride $s$, and kernel size $ks$ that can be approximated as:
$\#MAC \approx \frac{n}{s} \times oc \times ic \times ks$.
For the first convolution, with the SincConv's 8-bit activations and weights, a single \ac{mac} is equivalent to 64 bit-operations.
For the case of a small number of output channels \archparam{oc}${}=7$, \archparam{s}${}=20$, and $ks=51$, then the number of bit-operations is approximately 18 million.
For the same stride and kernel size, but with \archparam{oc}${}=14$, the number of bit-operations significantly increases to approximately 36 million.
Increasing \archparam{oc} raises the bit-operation cost of the SincConv significantly.
The \archparam{oc} parameter also determines the number of input channels for the first \ac{dsconv},
but it contributes to a smaller fraction of the total number of bit-operations.
Partly because the lower-precision 4-bit weights of the first \ac{dsconv} result in fewer bit-operations. 
Besides \archparam{oc} and \archparam{s}, the activation bit-width \archparam{a} also scales the bit-operation cost of the \ac{dsconv} blocks, since each \ac{mac} costs $4\times$\archparam{a} bit-operations under 4-bit weights.
\yu{consider moving the concrete examples (for the case of smaller.. larger oc1.. so so many bit operations) to a small table and refer to that instead, can be easier to follow.}

\subsubsection{Architectural Parameter Impact on Memory-Bits}
\label{sec:param_mem_bit}

Though the number of bit-operations varies widely, the overall model parameter count is mostly fixed, since the kernel size is fixed throughout the network.
The SincConv parameter count depends on the number of output channels \archparam{oc} and the kernel size $ks$ and is approximately $\#param_1 \approx \archparam{oc} \times ks$.
For the first \ac{dsconv}, the depthwise convolution has one filter per input channel, and the pointwise convolution has one filter per output channel.
Output channels are fixed at 160, as a middle ground among previous work's models that use \ac{dsconv}~\cite{helloegde:ARXIV2018}.
\yu{similar to previous section, consider moving concrete examples of configuration and memory bit impact into a table}
The parameter count is approximately $\#param_2 \approx \archparam{oc} \times 9 +  \archparam{oc} \times 160 = 169 \times \archparam{oc}$.
The rest of the \acp{dsconv} have a fixed parameter count, as does the final fully connected layer.
The stride does not affect the parameter count.
So, only the number of output channels of the SincConv \archparam{oc} affects it.
For \archparam{n_{ds}}${}=2$, the total parameter count is approximately $\#param \approx 220\,\archparam{oc} + 56{,}000$.
For \archparam{oc}${}=7$, the total parameter count is approximately $57{,}540$.
For \archparam{oc}${}=30$, the total parameter count is approximately $62{,}600$.
Changing \archparam{oc} therefore moves the total only a little relative to the large fixed \ac{dsconv} body; see \autoref{sec:results}.
Increasing \archparam{n_{ds}} adds a constant number of parameters per extra \ac{dsconv}.
\hms{Talks only about parameters while we have memory-bits in the title. But we don't vary the weights, so adding a statement could increase the focus on something we don't do.}

\subsection{Layer Tail Resource Cost}
\label{sec:layer_tail_resource_cost}

Batch normalization, quantization scaling, and quantized activations that follow the convolution are commonly referred to as the ``layer tail'' or non-\ac{mac} operations. 
SIRA~\cite{sira:TRETS2026} and other low-precision studies~\cite{pikelpn:CVPR2024, karami:ISPASS2025} have shown that when matrix multiplications are accelerated and quantized to low bit-widths, element-wise operations and layer tails dominate the total resource footprint, latency, and energy. 
We note that this problem is particularly acute for small models used in TinyML deployments, such as \texttt{Tartar}, and is shown in our resource breakdown  (\autoref{tab:fpga_breakdown_with_delta} in \autoref{sec:results}).

One can implement the layer tail with the composite approach~\cite{sira:TRETS2026}. 
In this case, each non-\ac{mac} operation
is implemented as a standalone module operating on floating-point numbers or on sufficiently accurate fixed-point arithmetic. 
\autoref{tab:composite_thresholding} reports hardware cost for a floating-point composite layer tail.
The total cost is already expensive for a lightweight design, especially considering that there are multiple layer tails in a given model.
The cost is relatively constant regardless of the activation bit-width \archparam{a}.

\looseness -1
Instead of the composite approach, one can opt for thresholding.
Thresholding is a technique for quantizing the accumulated values of the convolution output~\cite{sira:TRETS2026, Umuroglu:ARXIV2018}. 
In a typical layer tail of a \ac{nn}, all non-\ac{mac} operations can be absorbed into thresholding.
Although \autoref{tab:composite_thresholding} shows an exponential dependency on $\archparam{a}$ for thresholding, thresholding remains clearly preferable to composite for $\archparam{a}$ under six bits. 
Previous work also shows that thresholding is cheaper than fixed-point composite layer tails at low activation bit-widths~\cite{sira:TRETS2026}.

In typical low-bit-width quantized networks, the first and last layers are kept at 8-bit activation. 
However, even a single 8-bit activation can turn a lightweight model into a resource-heavy implementation. 
Motivated by \autoref{tab:composite_thresholding}, \texttt{Tartar} keeps \archparam{a} low (\archparam{a}${}\leq 6$) and only uses thresholding.
The activation bit-width of the final fully connected layer is also kept at six bits for the same reason.

\subsubsection{Architectural Parameter Impact on the Layer Tail}
\label{sec:param_layer_tail}

\texttt{Tartar} parametrizes the activation bit-width \archparam{a} as it has a significant impact on the layer tail cost. 
Thresholding compares the input against sorted thresholds and adds one for each threshold it exceeds. 
Each channel $c$ has up to $K = 2^{\archparam{a}} - 1$ sorted thresholds.
%
%
Each threshold must match the accumulator bit-width $\alpha$ ($16$--$24$ bits), while unused steps are padded with unreachable values ($+\infty$)~\cite{sira:TRETS2026}.
Then, storing uncompressed thresholds requires $M = \alpha \cdot (2^{\archparam{a}} - 1)$ bits per channel. 
Even for \archparam{a}${}\le6$, a large storage bottleneck can be observed. 
To overcome this storage bottleneck, we introduce delta-compressed thresholding.

\begin{table}[t]
    \centering
    \caption{Resources of Composite and Thresholding}
    \label{tab:composite_thresholding}
    \begin{tabular}{@{}lcc@{\hspace{5pt}}c@{\hspace{5pt}}c@{}}
    \textbf{Module} & \textbf{Origin} & \textbf{LUT} & \textbf{DSP} & \textbf{BRAM} \\
    \toprule
    Composite             & \archparam{a} = any bits   & 1659 & 8 & 1   \\    Threshold             & \archparam{a} = 8 bits & 100  & 0 & 32  \\
    Threshold             & \archparam{a} = 6 bits & 104  & 0 & 18   \\
    Threshold             & \archparam{a} = 5 bits & 93   & 0 & 9   \\
    Threshold             & \archparam{a} = 4 bits & 63   & 0 & 2.5 \\
    \bottomrule
    \end{tabular}
\end{table}

\section{Delta-Compressed Thresholding}
\label{sec:delta_thresholding}




To overcome the thresholding bottleneck, we observe that in networks with quantized ReLU activations like \texttt{Tartar}, thresholds follow an almost-linear sequence: affine scaling is perturbed by integer rounding, yielding a linear sequence offset by a small accumulated error similar to Bresenham's line drawing algorithm in graphics~\cite{bresenham1965algorithm}.
Defining the nominal step size as $step = \min_{i} \Delta_i$, the second-order error vector $e_i = \Delta_i - step$ strictly satisfies $e_i \in \{0, 1\}$. 
Setting $e_0 = 0$, the full threshold sequence is losslessly encoded by: 
\textit{(i)}~an $\alpha$-bit $base = T_0$, 
\textit{(ii)}~an $\alpha$-bit $step$, 
\textit{(iii)}~a 1-bit error vector $e \in \{0, 1\}^K$, and 
\textit{(iv)}
an \archparam{a}-bit valid step $count = K \le 2^{\archparam{a}} - 1$. 
Reconstruction is then exact:
\begin{equation}
\label{eq:reconstruct}
    T_0 = base, \qquad T_i = T_{i-1} + step + e_i.
\end{equation}

Consider a concrete example with $a=3$ bits ($K=7$). 
For trained thresholds $\mathit{T} = [100, 125, 150, 176, 201, 226, 252]$, first differences are $\mathit{\Delta} = [25, 25, 26, 25, 25, 26]$. 
With $base=100$ and $step=\min(\mathit{\Delta})=25$, the error vector is $\mathit{e} = [0, 0, 0, 1, 0, 0, 1]$. 
Reconstruction reproduces the exact values: $T_1 = 100+25+0=125$, while $e_3=1$ yields $T_3 = 150+25+1=176$. 
This reduces per-channel storage from $7 \times 16 = 112$ bits to $16\,(base) + 16\,(step) + 7\,(\mathit{e}) + 3\,(count) = 42$ bits ($2.67\times$ reduction). 
For $a=4$ ($K=15$), storage drops from 240 to 51 bits ($4.7\times$), and for $a=6$ ($K=63$), from 1{,}008 to 101 bits ($10.0\times$).
We apply this compression technique to all layers in \texttt{Tartar}, except the first SincConv layer, which incorporates non-linear $\log(|x|+1)$ compression before quantization~\cite{sincconv:ICASSP2020} and is not suitable for delta compression. 

\subsection{Hardware Architecture}
\label{sec:delta_hardware}

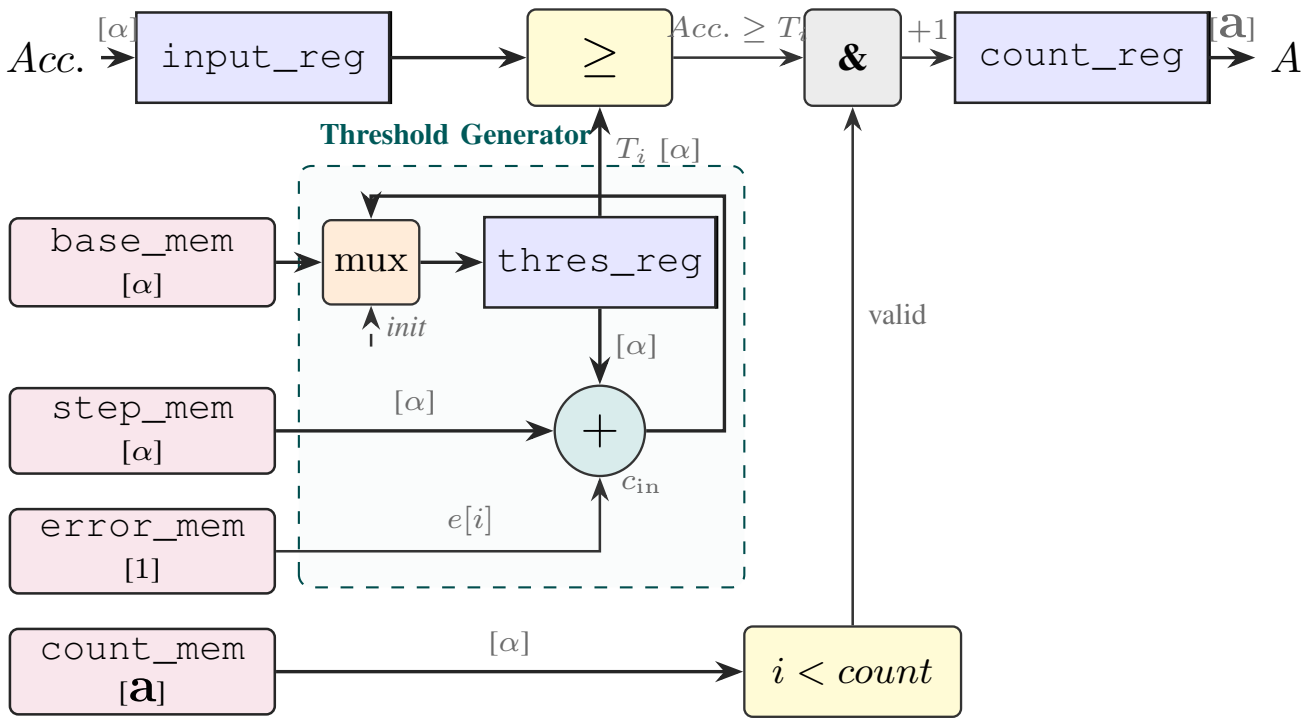
\begin{figure}[t]
    \centering
    \resizebox{\columnwidth}{!}{
\begin{tikzpicture}[
  font=\small,
  >=Stealth,
  every node/.style={align=center},
  reg/.style={
    draw=black!85,
    line width=0.6pt,
    fill=blue!10,
    sharp corners,
    minimum height=7.5mm,
    inner sep=3pt,
    append after command={
      \pgfextra
        \draw[line width=0.8pt] (\tikzlastnode.north east)
          -- (\tikzlastnode.south east);
      \endpgfextra
    },
  },
  mem/.style={
    draw=black!85,
    line width=0.6pt,
    fill=purple!10,
    rounded corners=2pt,
    minimum width=22mm,
    minimum height=7mm,
    inner sep=2pt,
  },
  fn/.style={
    draw=black!80,
    line width=0.55pt,
    rounded corners=2pt,
    inner sep=2pt,
  },
  wire/.style={draw=black!80, semithick},
  bus/.style={draw=black!85, thick},
  buslbl/.style={font=\scriptsize, text=gray!85!black},
]

\node[font=\normalsize] (in) at (-0.3, 3.8) {$Acc.$};

\node[reg, minimum width=21mm] (inreg) at (1.5, 3.8) {\texttt{input\_reg}};
\draw[bus, ->] (in) -- node[above, buslbl] {[$\alpha$]} (inreg.west);

\node[fn, fill=yellow!20, minimum width=12mm, minimum height=8mm] (comp) at (4.3, 3.8) {\large $\ge$};
\draw[bus, ->] (inreg.east) -- (comp.west);

\node[fn, fill=gray!15, minimum width=8mm, minimum height=8mm] (gate) at (6.4, 3.8) {\textbf{\&}};
\draw[wire, ->] (comp.east) -- node[above, buslbl] {$Acc. \ge T_i$} (gate.west);

\node[reg, minimum width=21mm] (cntreg) at (8.3, 3.8) {\texttt{count\_reg}};
\draw[wire, ->] (gate.east) -- node[above, buslbl] {$+1$} (cntreg.west);

\node[font=\normalsize] (out) at (10.0, 3.8) {$A$};
\draw[bus, ->] (cntreg.east) -- node[above, buslbl] {[$\archparam{a}$]} (out);

\node[mem] (basemem) at (0.5, 2.1) {\texttt{base\_mem}\\{\scriptsize [$\alpha$]}};
\node[mem] (stepmem) at (0.5, 0.7) {\texttt{step\_mem}\\{\scriptsize [$\alpha$]}};
\node[mem] (errmem)  at (0.5, -0.3) {\texttt{error\_mem}\\{\scriptsize [$1$]}};
\node[mem] (cntmem)  at (0.5, -1.3) {\texttt{count\_mem}\\{\scriptsize [$\archparam{a}$]}};

\draw[teal!60!black, dashed, rounded corners=3pt, line width=0.5pt, fill=teal!2] 
  (1.8, -0.6) rectangle (5.5, 2.9);
\node[font=\scriptsize\bfseries, text=teal!70!black, anchor=south west] 
  at (1.85, 2.95) {Threshold Generator};

\node[fn, fill=orange!15, minimum width=8mm, minimum height=7mm] (mux) at (2.4, 2.1) {$\mathrm{mux}$};
\draw[wire, dashed, ->] (2.4, 1.4) -- node[right, buslbl] {\textit{init}} (mux.south);

\node[reg, minimum width=19mm] (thresh) at (4.3, 2.1) {\texttt{thres\_reg}};

\node[fn, fill=teal!15, circle, inner sep=0pt, minimum size=7.5mm] (adder) at (4.3, 0.7) {\large $+$};
\node[font=\scriptsize, text=gray!80!black, anchor=south west] at (4.35, 0.06) {$c_\mathrm{in}$};

\draw[bus, ->] (basemem.east) -- (mux.west);

\draw[bus, ->] (mux.east) -- (thresh.west);

\draw[bus, ->] (thresh.north) -- node[pos=0.6, right, buslbl] {$T_i$ [$\alpha$]} (comp.south);

\draw[bus, ->] (thresh.south) -- node[right, buslbl] {[$\alpha$]} (adder.north);

\draw[bus, ->] (stepmem.east) -- node[above, buslbl] {[$\alpha$]} (adder.west);

\draw[wire, ->] (errmem.east) -| node[pos=0.3, above, buslbl] {$e[i]$} (adder.south);

\draw[bus, ->] (adder.east) -- ++(0.65, 0) |- (2.4, 2.65) -- (mux.north);

\node[fn, fill=yellow!20, minimum width=18mm, minimum height=7.5mm] (stepchk) at (6.4, -1.3) {$i < count$};
\draw[bus, ->] (cntmem.east) -- node[above, buslbl] {[$\alpha$]} (stepchk.west);
\draw[wire, ->] (stepchk.north) -- node[pos=0.6, right, buslbl] {valid} (gate.south);

\end{tikzpicture}}
    \caption{Architecture of the delta-compressed thresholding unit (single \ac{pe} datapath). Each cycle, the threshold accumulator increments by $step + e[i]$, where the 1-bit error acts directly as carry-in ($c_\mathrm{in}$). [\,] indicates bit-width.}
    \label{fig:delta_thresholding_hw}
\end{figure}

We translate this scheme into a Verilog hardware implementation, processing $C$ channels with a configurable number of \ac{pe} operating in parallel, as illustrated in \autoref{fig:delta_thresholding_hw}.
For simplicity, we assume that the variables $base$, $step$, and the current threshold $T$ are represented as $\alpha$-bit integers, although they could be implemented with different bit widths.
For each channel, the \ac{pe} maintains four memory arrays: \textit{base\_mem} ($A$ bits), \textit{step\_mem} ($\alpha$ bits), \textit{error\_mem} ($2^a - 1$ bits), and \textit{count\_mem} (\archparam{a} bits), requiring storage of:
\begin{equation}
\label{eq:new_mem_cost}
    M_\mathrm{delta} = 2\alpha + \archparam{a} + (2^{\archparam{a}} - 1) \quad \text{bits per channel,}
\end{equation}
removing the multiplicative factor $\alpha$ from the exponential ($2^{\archparam{a}}-1$) term. 
When input $x$ arrives, \textit{threshold\_reg} loads $base$, and \textit{count\_reg} resets to 0. 
Over $K = count$ cycles, the comparator evaluates $x \ge T_i$, incrementing \textit{count\_reg}. 
The next threshold is computed as $T_{i+1} = T_i + step + e[i+1]$. 
$e[i+1] \in \{0, 1\}$ is wired directly to the carry-in ($c_\mathrm{in}$) of the $\alpha$-bit adder, requiring no extra adder logic. 
The datapath per PE requires only one $\alpha$-bit adder with carry-in, one $\alpha$-bit comparator, and one $\alpha$-bit accumulator register.

\section{Methodology}
This section describes how we train \texttt{Tartar} (\autoref{sec:training_tartar}) and how we evaluate it on \ac{fpga} (\autoref{sec:evaluation}).

\subsection{Training \texttt{Tartar}}
\label{sec:training_tartar}

To train the \ac{pcm} model, we use Brevitas~\cite{brevitas:GITHUB2025} and extend it with the quantized sinc convolution layer. 
For all the layers, we use \ac{lsq} for learning the scales. 
The scales are initialized based on the statistics and then learned during training as in other \ac{qat} works~\cite{a2q:ICCV2023}.
Throughout the network, no bias is used.
The last layer is a fully connected layer with $12$ output channels using a per-tensor scale.
The other layers use a per-channel scale with no extra inference cost, as the scale can be absorbed into the batch normalization. 
The learning rate starts at $0.001$, and then it follows cosine annealing with warm restarts. 
We use a batch size of 256 and a dropout ratio of $0.1$.

We train on the \ac{gscd} v2~\cite{speechdataset:ARXIV2018}.
Following the standard 12-class setup of TinyML~\cite{mlperf:NEURIPS2021} for \ac{kws}, we use ten keywords, together with one silence and one unknown.
Remaining keywords are used for the unknown class.
For \ac{gscd} v2, this leaves 25 keywords to be used for the unknown class.
The silence class is created by using the background noise of the \ac{gscd}.
Validation and test sets are balanced across classes, while the training set is not.
Specifically, for v2, the training set has $\sim$54k samples for the unknown class, while each of the other classes has $\sim$3k samples. 

\subsection{Evaluation on FPGA}
\label{sec:evaluation}

We use FINN~\cite{finn-r:TRETS2018} to evaluate the resource requirements for \texttt{Tartar}.
FINN is a framework for mapping quantized neural networks to hardware, and provides a rich library of \ac{hls} and \ac{rtl} modules. 
FINN has two ways of parallelizing and folding the mapping, namely \ac{pe} and \ac{simd}.
\ac{pe} dictates how many output channels are worked on concurrently, while \ac{simd} dictates how many multiplies of weights and activations are done concurrently.

\texttt{Tartar} is implemented in a single QONNX file~\cite{qonnx:ARXIV2022}, but some of the layers are not supported by FINN~\cite{finn:FPGA2017}.
Specifically, the logarithm and the absolute value layers are not implemented.
The batch normalization is not quantized, which, together with the floating-point scaling factors, is aggregated into a single floating-point scale and bias~\cite{sira:TRETS2026}. 
After scale and bias aggregation and based on the range analysis, the accumulator bit-widths are minimized~\cite{sira:TRETS2026}, which results in reduced resource usage.
For the logarithm operator, as it is a monotonic function, it can be absorbed into the thresholds. 
The absolute value operator is not monotonic and cannot be absorbed into thresholds, and we implement it as a custom \ac{hls} module in FINN. 
We implement the delta-compression thresholding architecture, described in \autoref{sec:delta_hardware}, as a FINN \ac{rtl} module.

Performance is commonly a design target specified as the number of inferences per second (inf/s).
Previous works target throughputs of two, ten, and 25 inferences per second
(one detection every 500\,ms~\cite{fixedmfcc:AICAS2021}, 100\,ms~\cite{helloegde:ARXIV2018,data_reviewer_3:TCAD2020}, or 40\,ms~\cite{serialMFCC:JSCC2021}).
Two inferences per second is a practical lower bound for always-on \ac{kws}, even with overlapping windows.
FINN's auto-folding mechanism is used to calculate the lowest number of \ac{pe} and \ac{simd} required to meet the real-time requirements.
The main clock frequency of the design is also configurable, and it affects the throughput of the design. 

Post-implementation results are obtained for all configurations, targeting a Xilinx UltraScale+ \texttt{XCZU7EV}~\cite{zynqusp:DS891}.
This avoids resource congestion from penalizing models that otherwise would barely fit.
We then calculate the mean utilization based on the Artix-7 XC7A35T resources~\cite{artix7:DS180}, a device more suitable for edge deployment, as: 
$\frac{1}{3}\!\left(\frac{\mathrm{LUT}}{20800}+\frac{\mathrm{BRAM}}{50}+\frac{\mathrm{DSP}}{90}\right)$. 


\section{Results}
\label{sec:results}

The results cover the architectural parameters' impact on hardware resources and accuracy.

\subsection{Accuracy and Hardware Resources}
\label{sec:accuracy_exploration}


\newcount\AccVsArtixRow

\makeatletter
\newcommand{\FormatConfigName}[1]{\FormatConfigName@parse#1\@nil}
\def\FormatConfigName@parse oc#1-s#2-d#3-a#4-dc#5\@nil{%
  oc#1-s#2-n$_{ds}$#3-a#4%
}
\makeatother

\pgfplotsset{
  accvsartix/.style={
    scale only axis,
    width=7.2cm,
    height=5.4cm,
    grid=both,
    grid style={line width=.1pt, draw=gray!25},
    major grid style={line width=.2pt, draw=gray!45},
    xlabel={Mean Utilization},
    title style={font=\footnotesize},
    label style={font=\small},
    tick label style={font=\small},
    legend style={
      at={(0.98,0.02)},
      anchor=south east,
      font=\scriptsize,
      draw=none,
      fill=white,
      fill opacity=0.9,
    },
    xmin=0.00, xmax=0.6,
    ymin=91, ymax=96.5,
    xtick={0,0.1,0.2,0.3,0.4,0.5,0.6},
    ytick={91,92,93,94,95,96},
    enlargelimits=false,
    clip=true,
  },
  nodc/.style={only marks, mark=*, mark size=2.0pt, blue!70!black},
  dcmark/.style={only marks, mark=*, mark size=2.0pt, orange!85!black},
  pareto line/.style={draw=black!55, dashed, line width=0.8pt, mark=none},
}

\newcommand{\AccVsArtixClamp}[2]{%
  \pgfmathsetmacro{#1}{min(0.6,max(0.05,#2))}%
}
\newcommand{\AccVsArtixClampY}[2]{%
  \pgfmathsetmacro{#1}{min(96.5,max(91,#2))}%
}

\newcommand{\AccVsArtixPlot}[3]{%
  \pgfplotstableread[
    col sep=comma,
    header=false,
    skip first n=1,
  ]{#1}\accvsartixdata
  \gdef\AccVsArtixParetoCoords{}%
  \gdef\AccVsArtixNodcCoords{}%
  \gdef\AccVsArtixDcCoords{}%
  \pgfplotstablegetrowsof{\accvsartixdata}%
  \pgfmathtruncatemacro{\AccVsArtixN}{\pgfplotsretval}%
  \AccVsArtixRow=0
  \loop
    \ifnum\AccVsArtixRow<\AccVsArtixN
      \pgfplotstablegetelem{\the\AccVsArtixRow}{[index]7}\of\accvsartixdata
      \ifnum\pgfplotsretval=1\relax
        \pgfplotstablegetelem{\the\AccVsArtixRow}{[index]2}\of\accvsartixdata
        \AccVsArtixClamp{\AccU}{\pgfplotsretval}%
        \pgfplotstablegetelem{\the\AccVsArtixRow}{[index]3}\of\accvsartixdata
        \AccVsArtixClampY{\AccA}{\pgfplotsretval}%
        \pgfplotstablegetelem{\the\AccVsArtixRow}{[index]4}\of\accvsartixdata
        \edef\AccDc{\pgfplotsretval}%
        \pgfplotstablegetelem{\the\AccVsArtixRow}{[index]5}\of\accvsartixdata
        \edef\AccP{\pgfplotsretval}%
        \ifnum\AccP=1\relax
          \xdef\AccVsArtixParetoCoords{\AccVsArtixParetoCoords (\AccU,\AccA)}%
        \fi
        \ifnum\AccDc=0\relax
          \xdef\AccVsArtixNodcCoords{\AccVsArtixNodcCoords (\AccU,\AccA)}%
        \else
          \xdef\AccVsArtixDcCoords{\AccVsArtixDcCoords (\AccU,\AccA)}%
        \fi
      \fi
      \advance\AccVsArtixRow by 1
    \repeat
  \begin{tikzpicture}[trim axis left]
  \begin{axis}[accvsartix, ylabel={#3}]
    \addplot[pareto line] coordinates {\AccVsArtixParetoCoords};
    \addlegendentry{Pareto}
    \addplot[nodc] coordinates {\AccVsArtixNodcCoords};
    \addlegendentry{no delta-comp}
    \addplot[dcmark] coordinates {\AccVsArtixDcCoords};
    \addlegendentry{delta-comp}
    \addplot[
      only marks,
      mark=none,
      forget plot,
      point meta=explicit symbolic,
      nodes near coords={%
        \expandafter\FormatConfigName\expandafter{\pgfplotspointmeta}%
      },
      every node near coord/.style={
        font=\small,
        anchor=south west,
        inner sep=0.5pt,
        xshift=1.5pt,
        yshift=1.0pt,
      },
      x filter/.code={
        \pgfplotstablegetelem{\coordindex}{[index]6}\of\accvsartixdata
        \ifnum\pgfplotsretval=0\relax
          \def\pgfmathresult{}%
        \fi
      },
    ] table[
      col sep=comma,
      header=false,
      skip first n=1,
      x expr={min(0.6,max(0.05,\thisrowno{2}))},
      y expr={min(96.5,max(91,\thisrowno{3}))},
      meta index=1,
    ] {#1};
  \end{axis}
  \end{tikzpicture}%
}

\begin{figure*}[t]
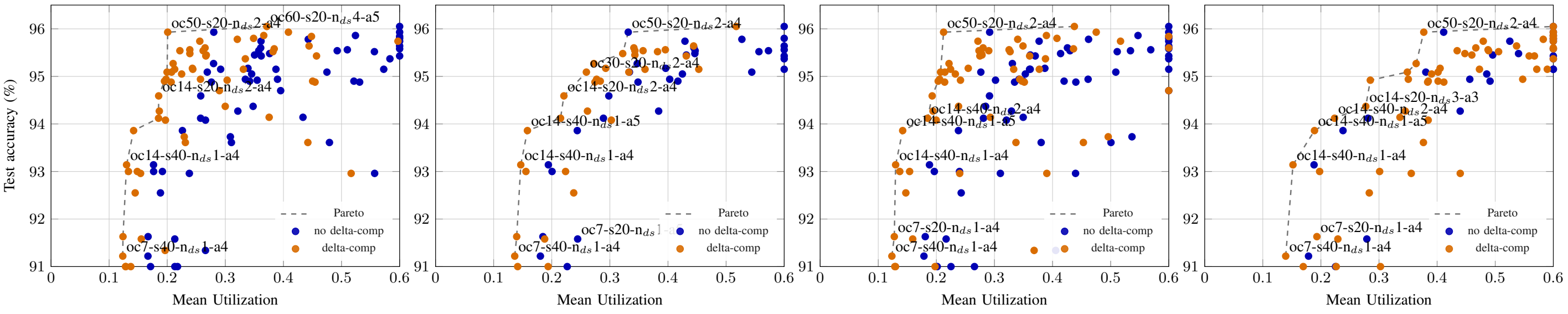

    \centering
    \begin{subfigure}[t]{0.24\textwidth}
        \centering
        \resizebox{\linewidth}{!}{%
          \AccVsArtixPlot%
            {graphs/acc_vs_artix_mhz10_fps2_ks51_both.csv}%
            {10\,MHz, 2\,inf/s}%
            {Test accuracy (\%)}%
        }
        \caption{2 inf/s at 10\,MHz}
        \label{fig:acc_vs_artix_mhz10_fps2}
    \end{subfigure}%
    \hfill
    \begin{subfigure}[t]{0.24\textwidth}
        \centering
        \resizebox{\linewidth}{!}{%
          \AccVsArtixPlot%
            {graphs/acc_vs_artix_mhz10_fps25_ks51_both.csv}%
            {10\,MHz, 25\,inf/s}%
            {}%
        }
        \caption{25 inf/s at 10\,MHz}
        \label{fig:acc_vs_artix_mhz10_fps25}
    \end{subfigure}%
    \hfill
    \begin{subfigure}[t]{0.24\textwidth}
        \centering
        \resizebox{\linewidth}{!}{%
          \AccVsArtixPlot%
            {graphs/acc_vs_artix_mhz5_fps2_ks51_both.csv}%
            {5\,MHz, 2\,inf/s}%
            {}%
        }
        \caption{2 inf/s at 5\,MHz}
        \label{fig:acc_vs_artix_mhz5_fps2}
    \end{subfigure}%
    \hfill
    \begin{subfigure}[t]{0.24\textwidth}
        \centering
        \resizebox{\linewidth}{!}{%
          \AccVsArtixPlot%
            {graphs/acc_vs_artix_mhz5_fps10_ks51_both.csv}%
            {5\,MHz, 10\,inf/s}%
            {}%
        }
        \caption{10 inf/s at 5\,MHz}
        \label{fig:acc_vs_artix_mhz5_fps10}
    \end{subfigure}
    \caption{Test accuracy versus mean Artix-7 XC7A35T utilization for \texttt{Tartar} models at a combination of frequencies and throughputs. Blue and orange points denote standard and delta-compressed thresholding, respectively. The dashed line is the Pareto front over both. Points on the plot borders could extend outside the plotted range and are outside our region of interest.}
    \label{fig:acc_vs_artix}
\end{figure*}


At 2\,inf/s, \ac{pe} and \ac{simd} are set to one, so the hardware resources are mostly influenced by the memory usage, as the required computational logic is the lowest possible.
As a result, increasing \archparam{oc} and decreasing \archparam{s} cost almost no additional hardware, because the design remains fully folded (see \autoref{sec:param_mem_bit}). 
The extra \ac{mac} operations are absorbed in time rather than in resources. 
For example, the \textit{configuration}: \archparam{oc}7-\archparam{s}40-\archparam{n_{ds}}1-\archparam{a}4 and the \textit{configuration}: \archparam{oc}14-\archparam{s}40-\archparam{n_{ds}}1-\archparam{a}4 
have similar resource usage, but a significant increase in accuracy is observed (see \autoref{fig:acc_vs_artix_mhz10_fps2}).
Opting for increasing \archparam{oc} and decreasing \archparam{s} is a good trade-off between accuracy and resource usage at low real-time requirements.

Auto-folding raises \ac{pe} and \ac{simd} above one to match higher throughputs, so computational logic begins to matter.
Configurations that have similar resource usage at 2\,inf/s diverge once their difference in bit-operation computational load forces different folding (see \autoref{sec:param_bitops}).
For example, \archparam{oc}7-\archparam{s}40-\archparam{n_{ds}}1-\archparam{a}4 and \archparam{oc}7-\archparam{s}20-\archparam{n_{ds}}1-\archparam{a}4 almost overlap in \autoref{fig:acc_vs_artix_mhz5_fps2}, yet they diverge in \autoref{fig:acc_vs_artix_mhz5_fps10}.
Larger \archparam{oc} and smaller \archparam{s} remain beneficial for accuracy, but are no longer a cheap resource trade-off once real-time constraints force extra hardware parallelism.

Pareto-optimal models are marked throughout the figures. 
For further exploration of the results, we keep three models of \autoref{fig:acc_vs_artix} as reference. 
The models are:
\newcommand{\tartarknob}[2]{\makebox[1.7cm][l]{\archparam{#1}${}=#2$}}%
\begin{itemize}
    \item \texttt{TartarS}: \tartarknob{oc}{7}\tartarknob{s}{40}\tartarknob{n_{ds}}{1}\tartarknob{a}{4}
    \item \texttt{TartarM}: \tartarknob{oc}{14}\tartarknob{s}{40}\tartarknob{n_{ds}}{2}\tartarknob{a}{4}
    \item \texttt{TartarL}: \tartarknob{oc}{50}\tartarknob{s}{20}\tartarknob{n_{ds}}{2}\tartarknob{a}{4}
\end{itemize}

All three reference models lie on the Pareto front of both \autoref{fig:acc_vs_artix_mhz10_fps2} (10\,MHz, 2\,inf/s) and \autoref{fig:acc_vs_artix_mhz5_fps10} (5\,MHz, 10\,inf/s).

\subsection{The Importance of the Layer Tail}
\label{sec:hw_exploration}

\begin{table}[t]
    \centering
    \caption{Post-synthesis resources breakdown of Composite}
    \label{tab:composite_only_results}
    \begin{tabular}{@{}lcc@{\hspace{5pt}}c@{\hspace{5pt}}c@{}}
    \textbf{Module} & \textbf{Origin} & \textbf{LUT} & \textbf{DSP} & \textbf{BRAM} \\
    \toprule
    Elementwise multiply  & BatchNorm      & 330 & 3 & 0.5  \\
    Elementwise add       & BatchNorm      & 343 & 2 & 0.5  \\
    Elementwise maximum   & ReLU           & 133 & 0 & 0    \\
    Elementwise multiply  & Quant Scale    & 164 & 3 & 0    \\
    Float-to-int          & Rounding       & 533 & 0 & 0    \\
    FIFO ($\times 4$)     & Extra Buffers  & 156 & 0 & 0    \\
    \hline
    Composite             & Total          & 1659 & 8 & 1   \\
    \bottomrule
    \end{tabular}
\end{table}

\looseness -1
For the resource cost of non-\ac{mac} computations, the activation bit-width is the key determinant. 
Increasing the thresholding bit width rapidly increases resource cost as explained in \autoref{sec:layer_tail_resource_cost} and shown in \autoref{tab:composite_thresholding}. 
Implementing the layer tail as a composite of floating-point operations uses fewer resources than thresholding at \archparam{a}${}=8$, but still consumes a significant amount for a lightweight implementation, especially in complex arithmetic as shown by the high DSP count. Even when using thresholding, the non-\ac{mac} operations remain relatively costly as seen in \autoref{tab:fpga_breakdown_with_delta}. 
Thresholding accounts for 59.3\% of all \ac{bram} in \texttt{TartarS} (8.0 out of 13.5), 57.1\% in \texttt{TartarM} (14.0 out of 24.5), and 58.2\% in \texttt{TartarL} (16.0 out of 27.5). 


Delta-compressed thresholding trades increased latency for reduced storage requirements.
At low throughputs, \texttt{Tartar} reduces resources using heavily folded \acp{mvau} ($PE=1, SIMD=1$), so \ac{mvau} tiles require hundreds to thousands of cycles per sample ($160$--$1{,}600$ cycles). 
This completely hides the $K \le 15$ (if \archparam{a}${}=4$) thresholding cycles behind \ac{mvau} compute as described in \autoref{sec:delta_thresholding}.
Because these cycles are fully hidden behind MVAU compute, delta-compressed thresholding introduces no end-to-end keyword detection latency penalty while yielding significant BRAM and LUT reductions (see \autoref{tab:fpga_breakdown_with_delta}).
Across evaluated clock and throughput targets, all but one Pareto-optimal point use delta-compressed thresholding.
The exception is \texttt{TartarL} at 10\,MHz and 25\,inf/s in \autoref{fig:acc_vs_artix_mhz10_fps25}, which suggests that for higher throughputs than what is required for \ac{kws}, the benefit of delta compression can diminish once folding requires increased parallelism.

\begin{table}[t]
\centering
\caption{Post-implementation FPGA resources with standard thresholding ($\mathrm{std}$) versus delta thresholding ($\Delta$). Each cell shows $\mathrm{std}\rightarrow\Delta$ at 10\,MHz and 2\,inf/s}
\label{tab:fpga_breakdown_with_delta}
\setlength{\tabcolsep}{3.5pt}
\begin{tabular}{@{}lccccc@{}}
\toprule
 &  & \multicolumn{2}{c}{\textbf{BRAM}} & \multicolumn{2}{c}{\textbf{LUT}} \\
\cmidrule(lr){3-4}\cmidrule(lr){5-6}
\textbf{Model} & \textbf{Acc. (\%)} & \textbf{\shortstack{Thresh.\\$\mathrm{std}\rightarrow\Delta$}} & \textbf{\shortstack{Total\\$\mathrm{std}\rightarrow\Delta$}} & \textbf{\shortstack{Thresh.\\$\mathrm{std}\rightarrow\Delta$}} & \textbf{\shortstack{Total\\$\mathrm{std}\rightarrow\Delta$}} \\
\midrule
\texttt{TartarS} & 91.22 & 8.0$\rightarrow$2.0 & 13.5$\rightarrow$7.5 & 626$\rightarrow$432 & 3{,}176$\rightarrow$2{,}969 \\
\texttt{TartarM} & 94.12 & 14.0$\rightarrow$3.5 & 24.5$\rightarrow$14 & 745$\rightarrow$605 & 4{,}028$\rightarrow$3{,}872 \\
\texttt{TartarL} & 95.93 & 16.0$\rightarrow$4.5 & 27.5$\rightarrow$16 & 732$\rightarrow$566 & 4{,}190$\rightarrow$4{,}028 \\
\bottomrule
\end{tabular}
\end{table}

\subsection{Comparison to Previous FPGA-based Works}
\label{sec:fpga_comparison}

\begin{table}[t]
\centering
\caption{FPGA: resource utilization and accuracy}
\label{tab:fpga_comparison}
\setlength{\tabcolsep}{3.2pt}
\begin{tabular}{@{}lcccccc@{}}
\textbf{Work} & \textbf{\shortstack{Raw\\PCM}} & \textbf{\shortstack{MFCC\\on FPGA}} & \textbf{BRAM} & \textbf{LUT} & \textbf{DSP} & \textbf{Acc. (\%)} \\
\midrule
Ng~\cite{smallfpgakws:ISCAS2024}        & No  & \textcolor{red}{NO}   & 20.5  & 16{,}000 & 0   & 91.01  \\
Dembeck~\cite{kwshls4ml:CAINE2024}      & No  & \textcolor{red}{NO}   & 53    & 25{,}762 & 93  & 91.48  \\
He~\cite{temporalkws:electronics2022}   & No  & \textcolor{red}{NO}   &  8    &  5{,}575 & 64  & 95.36  \\
Zhang~\cite{kwsbnn:entropy2025}         & No  & \textcolor{red}{NO}   & 6.5   &  1{,}939 & 0   & 97.29\textsuperscript{*}  \\
Ju~\cite{leekws:ICASSP2026}             & No  & \textcolor{teal}{YES}  & N/A   & 29{,}100 & 85  & 96.0   \\
Yang~\cite{directkws:IPMLP2024}         & \textcolor{teal}{YES}  & No  & 137   & 43{,}605 & 200 & 91.1   \\
\texttt{TartarS}                        & \textcolor{teal}{YES}  & No  & 7.5   &  2{,}969 & 7   & 91.22  \\
\texttt{TartarM}                        & \textcolor{teal}{YES}  & No  & 14    &  3{,}872 & 8   & 94.12  \\
\texttt{TartarL}                        & \textcolor{teal}{YES}  & No  & 16    &  4{,}028 & 8   & 95.93  \\
\midrule
\multicolumn{7}{l}{\textsuperscript{*}one keyword instead of the 12 keywords of TinyML} \\
\end{tabular}
\end{table}

Table~\ref{tab:fpga_comparison} shows recent FPGA keyword spotting implementations.
For \texttt{Tartar}, we report post-implementation results at 10\,MHz with one inference every 500\,ms. 
At 50\,MHz, all reference \texttt{Tartar} models achieve one inference every 40\,ms with the same resource utilization.

Ju et al.~\cite{leekws:ICASSP2026} introduce both a quantized model and an efficient hardware implementation of the \ac{mfcc}, but their resources are $7\times$ higher in \ac{lut} usage and over $10\times$ higher in \ac{dsp} usage than \texttt{TartarL}, which has similar accuracy.
The \ac{dsp} usage of their work barely fits on an Artix-7 XC7A35T (85 out of 90), while the \ac{lut} usage exceeds the device limit (29{,}100 out of 20{,}800).  
They report a memory usage of 17.9 KiB, but they do not report how many \acp{bram} are used to supply the required bandwidth for their model.
Yang et al.~\cite{directkws:IPMLP2024} introduce an encoder-decoder architecture operating directly on the \ac{pcm} waveform that learns how to behave like the \ac{mfcc} preprocessing, but this implementation is very resource-demanding. 

The remaining works in \autoref{tab:fpga_comparison} do not include the resources required for performing the \ac{mfcc}.
Even under such an unfair comparison, \texttt{Tartar} models remain competitive.

\section{Related Work}
\label{sec:related_work}

Mittermaier et al. with Sinc-Convolutions~\cite{sincconv:ICASSP2020} were the first to apply SincNet~\cite{sincnet:SLT2018} in the keyword spotting domain in combination with \ac{dsconv}.
Peter et al.~\cite{brevitassinc:ICASSP2022} quantize only the later layers and leave the heavy SincConv in floating point. 
\texttt{Tartar} quantizes the SincConv and enables an end-to-end quantized model for lightweight hardware deployment.
\texttt{TartarM} matches this accuracy class with 59.1k fully quantized parameters suitable for logic mapping.

Lee et al.~\cite{kws:JSSC2026} implement a fully integrated end-to-end \ac{asic} with sinc-based convolutions, though they target 
on-chip training and personalization.
Without on-chip training, they report an accuracy of 75.3\% on 35-keyword \ac{gscd} v2.
We retrain a small \texttt{Tartar} configuration (\archparam{oc}14-\archparam{s}20-\archparam{n_{ds}}1-\archparam{a}6)
that achieves 91.19\% on 35-keyword \ac{gscd} v2.
This exceeds their GSCD accuracy without personalization silicon, while their on-chip training targets accent adaptation on \ac{ascd}.

Several works use \ac{mfcc}~\cite{shan_aad_kws_jssc2023,li_mfcc_kws_jssc2024,lin_tsetlinkws_2025,leekws:ICASSP2026,data_reviewer_3:TCAD2020}, despite evidence that the front-end can consume half the latency or twice the network power~\cite{fixedmfcc:AICAS2021,serialMFCC:JSCC2021}.
Other works use learnable audio features in the analog domain, which brings other challenges, e.g., analog circuits are commonly sensitive to noise~\cite{cerutti_sub-mw_2022,hu_late_2024,hu_learnafe_2025,zhou_fwr_aicas2023,ray_kinget_jssc2023,shen_sensors_afe_2025,peaf_bergsma_2021,data_reviewer_4:nature2023}. 
Yoon et al.~\cite{coolbinary:COOLCHIPS2023} use a reusable BCNN inference engine that processes layers block-by-block under \ac{mcu} control, rather than a fully spatially mapped dataflow as in \texttt{Tartar}.

\section{Conclusion}
\label{sec:conclusion}

We propose \texttt{Tartar}, a keyword spotting architecture that operates on \textit{raw} waveforms and is designed for efficient mapping to hardware. 
Our \ac{fpga} evaluation shows that \texttt{Tartar}'s implementations are among the most lightweight while remaining competitive across accuracy targets. 
As future work, we plan to map \texttt{Tartar} to an \ac{asic} technology node and to adapt it to other audio AI applications. 

\bibliographystyle{IEEEtran}
\bibliography{refs,other}

@IEEEtranBSTCTL{IEEEtran:BSTcontrol,
  CTLuse_url = "no",
  CTLuse_forced_etal = "yes",
  CTLmax_names_forced_etal = "3",
  CTLnames_show_etal = "1",
  CTLdash_repeated_names = "no"
}

@Article{cerutti_sub-mw_2022,
    title = {Sub-{mW} Keyword Spotting on an {MCU}: Analog Binary Feature Extraction and Binary Neural Networks},
    pages = {2002-2012},
    doi = {10.1109/TCSI.2022.3142525},
    volume = {69},
    issue = {5},
    url = {https://ieeexplore.ieee.org/document/9707596},
    journal = TCAS,
    year = 2022,
    month = may,
    author = {Cerutti, Gianmarco and Cavigelli, Lukas and Andri, Renzo and Magno, Michele and Farella, Elisabetta and Benini, Luca},
}

@InProceedings{hu_late_2024,
    title = {Late Breaking Results: Circuit-Algorithm Co-design for Learnable Audio Analog Front-End},
    pages = {1-2},
    doi = {10.1145/3649329.3663496},
    url = {https://dl.acm.org/doi/10.1145/3649329.3663496},
    booktitle = DAC,
    year = 2024,
    month = jun,
    author = {Hu, Jinhai and Zhang, Zhongyi and Leow, Cong Sheng and Goh, Wang Ling and Gao, Yuan},
}

@Misc{lin_tsetlinkws_2025,
    title = {{TsetlinKWS}: A 65nm 16.58\,{$\mu$}{W}, 0.63mm$^2$ State-Driven Convolutional Tsetlin Machine-Based Accelerator for Keyword Spotting},
    doi = {10.48550/arXiv.2510.24282},
    url = {http://arxiv.org/abs/2510.24282},
    eprint = {arXiv:2510.24282},
    year = 2025,
    month = oct,
    author = {Lin, Baizhou and Fang, Yuetong and Xu, Renjing and Shafik, Rishad and Chauhan, Jagmohan},
}

@Misc{zynqusp:DS891,
    title = {Zynq {UltraScale+} {MPSoC} Data Sheet: Overview ({DS891})},
    url = {https://docs.amd.com/v/u/en-US/ds891-zynq-ultrascale-plus-overview},
    year = 2025,
    author = {{AMD}/{Xilinx}},
    note = {Includes {ZU7EV}/{XCZU7EV} device resources},
}

@Misc{artix7:DS180,
    title = {7 Series {FPGAs} Data Sheet: Overview ({DS180})},
    url = {https://docs.amd.com/v/u/en-US/ds180_7Series_Overview},
    year = 2020,
    author = {{AMD}/{Xilinx}},
    note = {Includes {Artix-7} {XC7A35T} device resources},
}

@InProceedings{pikelpn:CVPR2024,
    title = {{PikeLPN}: Mitigating Overlooked Inefficiencies of Low-Precision Neural Networks},
    pages = {15996-16005},
    booktitle = CVPR,
    year = 2024,
    author = {Neseem, Marina and McCullough, Conor and Hsin, Randy and Leichner, Chas and Li, Shan and Chong, In Suk and Howard, Andrew and Lew, Lukasz and Reda, Sherief and Rautio, Ville-Mikko and others},
}

@InProceedings{karami:ISPASS2025,
    title = {Understanding the Performance Horizon of the Latest {ML} Workloads with {NonGEMM} Workloads},
    pages = {1-14},
    booktitle = ISPASS,
    year = 2025,
    author = {Karami, Rachid and Kao, Sheng-Chun and Kwon, Hyoukjun},
}

@InProceedings{zhou_fwr_aicas2023,
    title = {High-Accuracy and Energy-Efficient Acoustic Inference using Hardware-Aware Training and a 0.34\,{nW}/{Ch} Full-Wave Rectifier},
    pages = {1-5},
    doi = {10.1109/AICAS57966.2023.10168561},
    url = {https://doi.org/10.1109/AICAS57966.2023.10168561},
    booktitle = AICAS,
    year = 2023,
    author = {Zhou, Sheng and Chen, Xi and Kim, Kwantae and Liu, Shih-Chii},
}

@Misc{hu_learnafe_2025,
    title = {{Learn{AFE}}: Circuit-Algorithm Co-design Framework for Learnable Audio Analog Front-End},
    doi = {10.48550/arXiv.2507.00755},
    url = {https://arxiv.org/abs/2507.00755},
    eprint = {arXiv:2507.00755},
    year = 2025,
    author = {Hu, Jinhai and Zhang, Zhongyi and Leow, Cong Sheng and Goh, Wang Ling and Gao, Yuan},
}

@Article{ray_kinget_jssc2023,
    title = {Ultra-Low-Power and Compact-Area Analog Audio Feature Extraction Based on Time-Mode Analog Filterbank Interpolation and Time-Mode Analog Rectification},
    pages = {919-932},
    doi = {10.1109/JSSC.2022.3227246},
    volume = {58},
    issue = {4},
    url = {https://doi.org/10.1109/JSSC.2022.3227246},
    journal = JSSC,
    year = 2023,
    author = {Ray, Subhajit and Kinget, Peter R.},
}

@Misc{peaf_bergsma_2021,
    title = {{PEAF}: Learnable Power Efficient Analog Acoustic Features for Audio Recognition},
    doi = {10.48550/arXiv.2110.03715},
    url = {https://arxiv.org/abs/2110.03715},
    eprint = {arXiv:2110.03715},
    year = 2021,
    month = oct,
    author = {Bergsma, Boris and Yang, Minhao and Cernak, Milos},
}

@Article{shen_sensors_afe_2025,
    title = {A {KWS} System for Edge-Computing Applications with Analog-Based Feature Extraction and Learned Step Size Quantized Classifier},
    pages = {2550},
    doi = {10.3390/s25082550},
    volume = {25},
    issue = {8},
    url = {https://doi.org/10.3390/s25082550},
    journal = {Sensors},
    year = 2025,
    author = {Shen, Yukai and Wu, Binyi and Straeussnigg, Dietmar and Gutierrez, Eric},
}

@Article{shan_aad_kws_jssc2023,
    title = {{AAD-KWS}: A Sub-{$\mu$}{W} Keyword Spotting Chip With an Acoustic Activity Detector Embedded in {MFCC} and a Tunable Detection Window in 28-nm {CMOS}},
    pages = {867-876},
    doi = {10.1109/JSSC.2022.3197838},
    volume = {58},
    issue = {3},
    url = {https://doi.org/10.1109/JSSC.2022.3197838},
    journal = JSSC,
    year = 2023,
    author = {Shan, Weiwei and Qian, Junyi and Zhu, Lixuan and Yang, Jun and Huang, Cheng and Cai, Hao},
}

@Article{li_mfcc_kws_jssc2024,
    title = {A 0.61-{$\mu$}{W} Fully Integrated Keyword-Spotting {ASIC} With Real-Point Serial {FFT}-Based {MFCC} and Temporal Depthwise Separable {CNN}},
    pages = {867-877},
    doi = {10.1109/JSSC.2023.3339528},
    volume = {59},
    issue = {3},
    url = {https://doi.org/10.1109/JSSC.2023.3339528},
    journal = JSSC,
    year = 2024,
    author = {Li, Cai and Zhi, Haochang and Yang, Kaiyue and Qian, Junyi and Yan, Zhihao and Zhu, Lixuan and Chen, Chao and Wang, Xi and Shan, Weiwei},
}

@Article{bresenham1965algorithm,
  author    = {Bresenham, Jack E.},
  title     = {Algorithm for computer control of a digital plotter},
  journal   = {IBM Systems Journal},
  volume    = {4},
  number    = {1},
  pages     = {25--30},
  year      = {1965},
  publisher = {IBM},
  doi       = {10.1147/sj.41.0025}
}

@STRING{ACM = {Association for Computing Machinery}}

@STRING{AFIPS = {Proceedings of the Spring Joint Computer Conference}}

@STRING{AICAS = {Proceedings of the {IEEE} International Conference on Artificial Intelligence Circuits and Systems}}

@STRING{CAINE = {Proceedings of the International Conference on Computer Appilcations in Industry and Engineering}}

@STRING{COMPUTER = {Computer}}

@STRING{COOLCHIPS = {Proceedings of the {IEEE} Symposium in Low-Power and High-Speed Chips}}

@STRING{CVPR = {Proceedings of the {IEEE}/{CVF} Conference on Computer Vision and Pattern Recognition}}

@STRING{DAC = {Proceedings of the {ACM}/{IEEE} Design Automation Conference}}

@STRING{FPGA = {Proceedings of the {ACM}/{SIGDA} International Symposium on Field-Programmable Gate Arrays}}

@STRING{ICASSP = {Proceedings of the {IEEE} International Conference on Acoustics, Speech and Signal Processing}}

@STRING{ICCV = {Proceedings of the {IEEE}/{CVF} International Conference on Computer Vision}}

@STRING{ICLR = {International Conference on Learning Representations}}

@STRING{IEEE = {IEEE Computer Society}}

@STRING{IPMLP = {Proceedings of the {ACM} International Conference on Image Processing, Machine Learning, and Pattern Recognition}}

@STRING{ISCAS = {Proceedings of the International Symposium on Circuits and Systems}}

@STRING{ISPASS = {Proceedings of the International Symposium on Performance Analysis of Systems and Software}}

@STRING{JSSC = {{IEEE} Journal of Solid-State Circuits}}

@STRING{NEURIPS = {Proceedings of the Conference on Neural Information Processing Systems}}

@STRING{SLT = {Proceedings of the {IEEE} Spoken Language Technology Workshop}}

@STRING{TCAS = {{IEEE} Transactions on Circuits and Systems I}}

@STRING{TOCS = {{ACM}  Transactions on Computer Systems}}

@STRING{TRETS = {{ACM} Transactions on Reconfigurable Technology and Systems}}

@Article{kws:JSSC2026,
    zotero-key = {78NVV7XJ},
    title = {A 7.5$\mu$W 35-Keyword End-to-End Keyword Spotting System With Random Augmented On-Chip Training},
    pages = {4321-4330},
    doi = {10.1109/JSSC.2026.3660634},
    volume = {61},
    issue = {8},
    url = {https://ieeexplore.ieee.org/document/11399878/},
    journal = JSSC,
    zotero-modified = {2026-09-18T16:31:31Z},
    year = 2026,
    month = aug,
    author = {Lee, Hyuk-Jin and Pyo, Kyunghoon and Jang, Taekwang and Seok, Mingoo and Cho, SeongHwan},
}

@Article{data_reviewer_3:TCAD2020,
    zotero-key = {59YPHX4J},
    title = {{UltraTrail}: A Configurable Ultralow-Power TC-ResNet AI Accelerator for Efficient Keyword Spotting},
    pages = {4240-4251},
    doi = {10.1109/TCAD.2020.3012320},
    volume = {39},
    issue = {11},
    url = {https://ieeexplore.ieee.org/document/9216480/},
    Journal = {IEEE Transactions on Computer-Aided Design of Integrated Circuits and Systems},
    zotero-modified = {2026-09-17T17:47:19Z},
    year = 2020,
    month = nov,
    author = {Bernardo, Paul Palomero and Gerum, Christoph and Frischknecht, Adrian and Lübeck, Konstantin and Bringmann, Oliver},
}

@Article{data_reviewer_4:nature2023,
    zotero-key = {W9SAXDTD},
    title = {An analog-AI chip for energy-efficient speech recognition and transcription},
    publisher = {Nature Publishing Group},
    pages = {768-775},
    doi = {10.1038/s41586-023-06337-5},
    volume = {620},
    issue = {7975},
    url = {https://www.nature.com/articles/s41586-023-06337-5},
    Journal = {Nature},
    zotero-modified = {2026-09-17T16:33:47Z},
    year = 2023,
    month = aug,
    author = {Ambrogio, S. and Narayanan, P. and Okazaki, A. and Fasoli, A. and Mackin, C. and Hosokawa, K. and Nomura, A. and Yasuda, T. and Chen, A. and Friz, A. and Ishii, M. and Luquin, J. and Kohda, Y. and Saulnier, N. and Brew, K. and Choi, S. and Ok, I. and Philip, T. and Chan, V. and Silvestre, C. and Ahsan, I. and Narayanan, V. and Tsai, H. and Burr, G. W.},
}

@InProceedings{mobilenet_bad_quant:CVPR2021,
    zotero-key = {QR2JAAZN},
    title = {Do All {MobileNets} Quantize Poorly? Gaining Insights Into the Effect of Quantization on Depthwise Separable Convolutional Networks Through the Eyes of Multi-Scale Distributional Dynamics},
    doi = {10.48550/arXiv.2104.11849},
    url = {https://doi.org/10.48550/arXiv.2104.11849},
    booktitle = CVPR,
    zotero-modified = {2026-08-05T08:57:03Z},
    year = 2021,
    month = jun,
    author = {Yun, Stone and Wong, Alexander},
}

@Article{serialMFCC:JSCC2021,
    zotero-key = {ED8ZK6AN},
    title = {A 510-{nW} Wake-Up Keyword-Spotting Chip Using Serial-{FFT}-Based {MFCC} and Binarized Depthwise Separable {CNN} in 28-nm CMOS},
    pages = {151-164},
    doi = {10.1109/JSSC.2020.3029097},
    volume = {56},
    issue = {1},
    url = {https://ieeexplore.ieee.org/document/9233931},
    journal = JSSC,
    zotero-modified = {2026-08-05T08:56:56Z},
    year = 2021,
    month = jan,
    author = {Shan, Weiwei and Yang, Minhao and Wang, Tao and Lu, Yicheng and Cai, Hao and Zhu, Lixuan and Xu, Jiaming and Wu, Chengjun and Shi, Longxing and Yang, Jun},
}

@InProceedings{leaf:ICLR2021,
    zotero-key = {UFFK7ZS2},
    title = {{LEAF}: A Learnable Frontend for Audio Classification},
    doi = {10.48550/arXiv.2101.08596},
    url = {http://arxiv.org/abs/2101.08596},
    booktitle = ICLR,
    zotero-modified = {2026-08-05T08:56:56Z},
    year = 2021,
    month = jan,
    author = {Zeghidour, Neil and Teboul, Olivier and Quitry, Félix de Chaumont and Tagliasacchi, Marco},
}

@Misc{mobilenet_v1:ARXIV2017,
    zotero-key = {CVLDEQW5},
    title = {{MobileNets}: Efficient Convolutional Neural Networks for Mobile Vision Applications},
    doi = {10.48550/arXiv.1704.04861},
    url = {http://arxiv.org/abs/1704.04861},
    eprint = {arXiv:1704.04861},
    zotero-modified = {2026-08-05T08:56:55Z},
    year = 2017,
    month = apr,
    author = {Howard, Andrew G. and Zhu, Menglong and Chen, Bo and Kalenichenko, Dmitry and Wang, Weijun and Weyand, Tobias and Andreetto, Marco and Adam, Hartwig},
}

@Article{kwsbnn:entropy2025,
    zotero-key = {EIFY6HI7},
    title = {High Precision Speech Keyword Spotting Based on Binary Deep Neural Network in {FPGA}},
    publisher = {Multidisciplinary Digital Publishing Institute},
    pages = {1143},
    doi = {10.3390/e27111143},
    volume = {27},
    issue = {11},
    url = {https://www.mdpi.com/1099-4300/27/11/1143},
    Journal = {Entropy},
    zotero-modified = {2026-08-05T08:56:53Z},
    year = 2025,
    author = {Zhang, Ang and Shi, Jialiang and Qian, Hui and Wang, Junjie},
}

@Article{bitoperation:TOCS2021,
    zotero-key = {IMEEU5AE},
    title = {{UNIQ}: Uniform Noise Injection for Non-Uniform Quantization of Neural Networks},
    publisher = ACM,
    pages = {4:1–4:15},
    doi = {10.1145/3444943},
    volume = {37},
    issue = {1-4},
    url = {https://dl.acm.org/doi/10.1145/3444943},
    journal = TOCS,
    zotero-modified = {2026-08-05T08:56:52Z},
    year = 2021,
    month = mar,
    author = {Baskin, Chaim and Liss, Natan and Schwartz, Eli and Zheltonozhskii, Evgenii and Giryes, Raja and Bronstein, Alex M. and Mendelson, Avi},
}

@InProceedings{amdahl:AFIPS1967,
    zotero-key = {FWDD3WXW},
    title = {Validity of the single processor approach to achieving large scale computing capabilities},
    pages = {483–485},
    doi = {10.1145/1465482.1465560},
    url = {https://dl.acm.org/doi/10.1145/1465482.1465560},
    booktitle = AFIPS,
    zotero-modified = {2026-07-17T15:31:44Z},
    year = 1967,
    month = apr,
    author = {Amdahl, Gene M.},
}

@InProceedings{mlperf:NEURIPS2021,
    zotero-key = {GWRUPDYZ},
    title = {{MLPerf} Tiny Benchmark},
    doi = {10.48550/arXiv.2106.07597},
    url = {https://doi.org/10.48550/arXiv.2106.07597},
    booktitle = NEURIPS,
    zotero-modified = {2026-07-16T23:32:13Z},
    year = 2021,
    month = dec,
    author = {Banbury, Colby and Reddi, Vijay Janapa and Torelli, Peter and Holleman, Jeremy and Jeffries, Nat and Kiraly, Csaba and Montino, Pietro and Kanter, David and Ahmed, Sebastian and Pau, Danilo and Thakker, Urmish and Torrini, Antonio and Warden, Peter and Cordaro, Jay and Guglielmo, Giuseppe Di and Duarte, Javier and Gibellini, Stephen and Parekh, Videet and Tran, Honson and Tran, Nhan and Wenxu, Niu and Xuesong, Xu},
}

@Misc{speechdataset:ARXIV2018,
    zotero-key = {FGXEEDLP},
    title = {Speech Commands: A Dataset for Limited-Vocabulary Speech Recognition},
    doi = {10.48550/arXiv.1804.03209},
    url = {http://arxiv.org/abs/1804.03209},
    eprint = {arXiv:1804.03209},
    zotero-modified = {2026-07-16T22:18:56Z},
    year = 2018,
    month = apr,
    author = {Warden, Pete},
}

@Article{temporalkws:electronics2022,
    zotero-key = {D7ESFHHZ},
    title = {A Configurable Accelerator for Keyword Spotting Based on Small-Footprint Temporal Efficient Neural Network},
    publisher = {Multidisciplinary Digital Publishing Institute},
    pages = {2571},
    doi = {10.3390/electronics11162571},
    volume = {11},
    issue = {16},
    url = {https://www.mdpi.com/2079-9292/11/16/2571},
    Journal = {Electronics},
    zotero-modified = {2026-07-14T22:52:39Z},
    year = 2022,
    author = {He, Keyan and Chen, Dihu and Su, Tao},
}

@InProceedings{kwshls4ml:CAINE2024,
    zotero-key = {L98CGXQX},
    title = {Development and Optimization of an Ultra-lightweight Deep Spoken Keyword Spotting Model for {FPGA} Acceleration},
    pages = {3-20},
    doi = {10.1007/978-3-031-76273-4_1},
    booktitle = CAINE,
    zotero-modified = {2026-07-11T17:04:41Z},
    year = 2025,
    author = {Dembeck, Trysten and Parikh, Chirag},
    editor = {Hu, Gongzhu and Kambhampaty, Krishna K. and Roy, Indranil},
}

@InProceedings{coolbinary:COOLCHIPS2023,
    zotero-key = {2B755YYJ},
    title = {A Real-Time Keyword Spotting System Based on an End-To-End Binary Convolutional Neural Network in {FPGA}},
    pages = {1-3},
    doi = {10.1109/COOLCHIPS57690.2023.10121981},
    url = {https://ieeexplore.ieee.org/document/10121981},
    booktitle = COOLCHIPS,
    zotero-modified = {2026-07-03T16:55:35Z},
    year = 2023,
    month = apr,
    author = {Yoon, Jinsung and Lee, Donghyun and Kim, Neungyun and Lee, Su-Jung and Kwak, Gil-Ho and Kim, Tae-Hwan},
}

@InProceedings{fixedmfcc:AICAS2021,
    zotero-key = {9PL4HHNM},
    title = {Integer-Only Approximated {MFCC} for Ultra-Low Power Audio {NN} Processing on Multi-Core {MCU}s},
    pages = {1-4},
    doi = {10.1109/AICAS51828.2021.9458491},
    url = {https://ieeexplore.ieee.org/document/9458491},
    booktitle = AICAS,
    zotero-modified = {2026-07-03T16:55:14Z},
    year = 2021,
    month = jun,
    author = {Fariselli, Marco and Rusci, Manuele and Cambonie, Joel and Flamand, Eric},
}

@InProceedings{yadav:INTERSPEECH2022,
    zotero-key = {CKNIHF9N},
    title = {Learning neural audio features without supervision},
    url = {https://arxiv.org/abs/2203.15519v1},
    booktitle = {Interspeech},
    zotero-modified = {2026-07-01T17:05:41Z},
    year = 2022,
    author = {Yadav, Sarthak and Zeghidour, Neil},
}

@InProceedings{smallfpgakws:ISCAS2024,
    zotero-key = {7X67B4I8},
    title = {High Accuracy and Low Latency Mixed Precision Neural Network Acceleration for {TinyML} Applications on Resource-Constrained {FPGAs}},
    pages = {1-5},
    doi = {10.1109/ISCAS58744.2024.10558440},
    url = {https://ieeexplore.ieee.org/document/10558440},
    booktitle = ISCAS,
    zotero-modified = {2026-07-01T12:59:28Z},
    year = 2024,
    month = may,
    author = {Ng, Wei Soon and Ling Goh, Wang and Gao, Yuan},
}

@InProceedings{sincconv:ICASSP2020,
    zotero-key = {JP44FAZE},
    title = {Small-Footprint Keyword Spotting on Raw Audio Data with Sinc-Convolutions},
    pages = {7454-7458},
    doi = {10.1109/ICASSP40776.2020.9053395},
    url = {https://ieeexplore.ieee.org/document/9053395},
    booktitle = ICASSP,
    zotero-modified = {2026-06-26T00:11:46Z},
    year = 2020,
    month = may,
    author = {Mittermaier, Simon and Kürzinger, Ludwig and Waschneck, Bernd and Rigoll, Gerhard},
}

@InProceedings{brevitassinc:ICASSP2022,
    zotero-key = {6XFPZE3V},
    title = {End-to-end Keyword Spotting using Neural Architecture Search and Quantization},
    doi = {10.1109/ICASSP43922.2022.9746535},
    url = {https://doi.org/10.1109/ICASSP43922.2022.9746535},
    booktitle = ICASSP,
    zotero-modified = {2026-06-26T00:11:41Z},
    year = 2021,
    month = apr,
    author = {Peter, David and Roth, Wolfgang and Pernkopf, Franz},
}

@InProceedings{directkws:IPMLP2024,
    zotero-key = {YM4CCV3Z},
    title = {An {FPGA}-Accelerated Approach to Direct Waveform Processing for Low-Power Keyword Spotting},
    pages = {157–161},
    doi = {10.1145/3700906.3700931},
    url = {https://dl.acm.org/doi/10.1145/3700906.3700931},
    booktitle = IPMLP,
    zotero-modified = {2026-06-26T00:11:16Z},
    year = 2024,
    month = dec,
    author = {Yang, Yucheng and Duan, Shukai and Wang, Lidan},
}

@Misc{helloegde:ARXIV2018,
    zotero-key = {UVTS6NQ8},
    title = {Hello Edge: Keyword Spotting on Microcontrollers},
    doi = {10.48550/arXiv.1711.07128},
    url = {http://arxiv.org/abs/1711.07128},
    eprint = {arXiv:1711.07128},
    zotero-modified = {2026-06-26T00:11:13Z},
    year = 2018,
    month = feb,
    author = {Zhang, Yundong and Suda, Naveen and Lai, Liangzhen and Chandra, Vikas},
}

@InProceedings{leekws:ICASSP2026,
    zotero-key = {7QQB6SPD},
    title = {{FPGA} Implementation of Accurate and Low-Cost Keyword Spotting},
    pages = {18762-18766},
    doi = {10.1109/ICASSP55912.2026.11463209},
    url = {https://ieeexplore.ieee.org/document/11463209},
    booktitle = ICASSP,
    zotero-modified = {2026-06-23T12:55:30Z},
    year = 2026,
    month = may,
    author = {Ju, Sanghyeon and Lee, Sunggu},
}

@InProceedings{sincnet:SLT2018,
    zotero-key = {XMRYXP3J},
    title = {Speaker Recognition from Raw Waveform with {SincNet}},
    doi = {10.1109/SLT.2018.8639585},
    url = {https://doi.org/10.1109/SLT.2018.8639585},
    booktitle = SLT,
    zotero-modified = {2026-06-14T22:14:09Z},
    year = 2019,
    month = aug,
    author = {Ravanelli, Mirco and Bengio, Yoshua},
}

@Article{sira:TRETS2026,
    zotero-key = {3J49VAYK},
    title = {{SIRA}: Scaled-Integer Range Analysis for Optimizing {FPGA} Dataflow Neural Network Accelerators},
    publisher = {ACM Transactions on Reconfigurable Technology and Systems (TRETS)},
    doi = {10.1145/3807510},
    url = {https://doi.org/10.1145/3807510},
    journal = TRETS,
    zotero-modified = {2026-05-13T09:44:58Z},
    year = 2026,
    month = apr,
    author = {Umuroglu, Yaman and Berganski, Christoph and Jentzsch, Felix and Danilowicz, Michal and Kryjak, Tomasz and Bezaitis, Charalampos and Sjalander, Magnus and Colbert, Ian and Preusser, Thomas and Petri-Koenig, Jakoba and Blott, Michaela},
}

@InProceedings{a2q:ICCV2023,
    zotero-key = {SUGFSTIQ},
    title = {{A2Q}: Accumulator-Aware Quantization with Guaranteed Overflow Avoidance},
    pages = {16943-16952},
    doi = {10.1109/ICCV51070.2023.01558},
    url = {https://doi.org/10.1109/ICCV51070.2023.01558},
    booktitle = ICCV,
    zotero-modified = {2026-01-13T15:18:49Z},
    year = 2023,
    author = {Colbert, Ian and Pappalardo, Alessandro and Petri-Koenig, Jakoba},
}

@Misc{brevitas:GITHUB2025,
    zotero-key = {I6HQ63PT},
    title = {Xilinx/brevitas: Release v0.12.1},
    url = {https://zenodo.org/records/16987789},
    zotero-modified = {2025-09-07T16:47:45Z},
    year = 2025,
    author = {Pappalardo, Alessandro and Franco, Giuseppe and nickfraser and Colbert, Ian and Lago, Pablo Monteagudo and Grob, Fabian and Costigan, Timothy and Savolainen, Oscar and Stoian, Andrei and jinchen and Gerdelan, Anton and alexredd99 and Jentzsch, Felix and Duarte, Javier and Montero, Luis and MichalMachura and Peracha, Omar and Khan, Saad and Witham, Kenneth L. and Paine, Tim and Umuroglu, Yaman and derpda and hkayann and vfdev and rgb000000},
}

@InProceedings{finn:FPGA2017,
    zotero-key = {NE8U5V77},
    title = {{FINN}: A Framework for Fast, Scalable Binarized Neural Network Inference},
    pages = {65-74},
    doi = {10.1145/3020078.3021744},
    url = {https://doi.org/10.1145/3020078.3021744},
    booktitle = FPGA,
    zotero-modified = {2025-05-13T07:53:44Z},
    year = 2017,
    month = feb,
    author = {Umuroglu, Yaman and Fraser, Nicholas J. and Gambardella, Giulio and Blott, Michaela and Leong, Philip and Jahre, Magnus and Vissers, Kees},
}

@Article{finn-r:TRETS2018,
    zotero-key = {UDIKKMUA},
    title = {{FINN}-{R}: An end-to-end deep-learning framework for fast exploration of quantized neural networks},
    pages = {1–23},
    doi = {10.1145/3242897},
    volume = {11},
    issue = {3},
    url = {https://doi.org/10.1145/3242897},
    journal = TRETS,
    zotero-modified = {2025-05-13T07:53:43Z},
    year = 2018,
    author = {Blott, Michaela and Preußer, Thomas B. and Fraser, Nicholas J. and Gambardella, Giulio and O’brien, Kenneth and Umuroglu, Yaman and Leeser, Miriam and Vissers, Kees},
}

@Misc{Umuroglu:ARXIV2018,
    zotero-key = {KDGBZU9J},
    title = {Streamlined Deployment for Quantized Neural Networks},
    url = {http://arxiv.org/abs/1709.04060},
    archivePrefix = {arXiv},
    eprint = {1709.04060},
    zotero-modified = {2023-06-01T20:15:41Z},
    year = 2018,
    month = may,
    author = {Umuroglu, Yaman and Jahre, Magnus},
}

@Misc{qonnx:ARXIV2022,
    zotero-key = {KNGM5H8U},
    title = {{QONNX}: Representing Arbitrary-Precision Quantized Neural Networks},
    url = {http://arxiv.org/abs/2206.07527},
    archivePrefix = {arXiv},
    eprint = {2206.07527},
    zotero-modified = {2023-06-01T20:15:39Z},
    year = 2022,
    month = jun,
    author = {Pappalardo, Alessandro and Umuroglu, Yaman and Blott, Michaela and Mitrevski, Jovan and Hawks, Ben and Tran, Nhan and Loncar, Vladimir and Summers, Sioni and Borras, Hendrik and Muhizi, Jules and Trahms, Matthew and Hsu, Shih-Chieh and Hauck, Scott and Duarte, Javier},
}



\end{document}